# Spin relaxation dynamics of radical-pair processes at low magnetic fields

T. H. Tennahewa,[1] S. Hosseinzadeh,[1] S. I. Atwood,[1] H. Popli,[1] H. Malissa,[1,2] J. M. Lupton,[1,2] and C. Boehme[1]

[1]Department of Physics and Astronomy, University of Utah, Salt Lake City, Utah 84112, USA

[2]Institut für Experimentelle und Angewandte Physik, Universität Regensburg, Universitätsstrasse 31, 93053 Regensburg, Germany

## Abstract

We report measurements of room-temperature spin-relaxation times $T_1$ and $T_2$ of charge-carrier spins in a π-conjugated polymer thin film under bipolar injection and low (1 mT ≲ $B_0$ ≲ 10 mT) static magnetic fields, using electrically detected magnetic resonant Hahn-echo and inversion-recovery pulse sequences. The experiments confirm the correlation between the magnetic-field sensitive observables of radical-pair processes, which include both the spin-dependent recombination currents in organic semiconductors and the associated spin-relaxation times when random local hyperfine fields and external magnetic fields compete in magnitude. Whereas a striking field dependence of spin-lattice relaxation exists in the low-field regime, the apparent spin decoherence time remains field independent as the distinction between the two is lifted at low fields.

The strong magnetic-field sensitivity of radical-pair systems has intrigued researchers for decades, ranging from biologists studying avian magneto-reception [1,2] and chemists exploring magnetic-field-dependent reaction rates [3,4], all the way to materials scientists and condensed matter physicists investigating the substantial low-field magnetoresistance of organic [5,6] and some disordered inorganic [7] semiconductors, where charge-carrier recombination currents follow radical-pair physics. Common to all these different systems is that they exhibit observables governed by non-equilibrium electronic relaxation processes that are subject to spin selection rules caused by weak spin-orbit coupling [8]. As a result of this research, it has been well established that the strong magnetic-field dependence of these processes is caused by a competition between externally applied magnetic fields and local internal hyperfine fields that are distributed randomly in magnitude and direction throughout a given radical-pair ensemble [9]. When external fields become dominant over local hyperfine fields, the paramagnetic electronic states experience a change of their individual spin quantization axis, which, in turn, changes the permutation symmetry $\{|S\rangle\langle S|, \tilde{\rho}\}^+$ of the radical-pair spin ensemble $\tilde{\rho}$ [10], which causes changes to the spin-dependent transition rate and, thus, to observables such as recombination current [11], light emission [12] or the perception of light by certain bird species [13]. While there is an abundance of data corroborating the magnetic-field dependence of observables governed by the radical-pair mechanism, there is still one key implication of this model that has not previously been scrutinized experimentally, namely the pronounced magnetic-field dependence of the longitudinal spin-relaxation time $T_1$, as predicted by the radical-pair model [2], and, possibly, also of the transverse spin-relaxation time $T_2$. The second effect should be pronounced within the applied magnetic-field range in which the radical pairs experience changes of the axis of quantization [5]. As the direction of a general quantization axis becomes ill-defined for many spin pairs within this range, the definition of longitudinal and transverse directions loses its meaning, along with the conventional definitions of $T_1$ and $T_2$, and the two observables become indistinguishable. The key challenge for an experimental test of this effect is to measure spin relaxation at almost vanishing magnetic fields. Spin-relaxation times are most accurately determined using pulsed magnetic resonance techniques; however, these methods typically require strong magnetic fields in order to generate sufficient spin polarization for the inductive detection of spin resonance, and, therefore, cannot be applied to the range of low magnetic fields of relevance here.

In this letter, we present experimental evidence for significant variations in the spin-relaxation rate of a radical-pair system in the presence of externally applied magnetic fields with magnitudes close to those of random hyperfine fields. We have overcome the above-mentioned experimental challenges by conducting pulsed electrically detected magnetic resonance (pEDMR) spectroscopy using the spin-dependent recombination current in a thin film of the commercial light-emitting conjugated polymer Superyellow-PPV (SY-PPV), a poly(phenylene-vinylene) copolymer (Sigma Aldrich) used for organic light-emitting diodes (OLEDs). These currents are governed by radical-pair processes [14], allowing for the observation of magnetic resonance without the need of spin polarization. The measurements can therefore be performed at arbitrarily low magnetic fields and thus spin-resonant frequencies.

The low-field magnetoconductance of a SY-PPV OLED (at a base current of $I \approx 20$ µA) is shown in Figure 1(a). The curve exhibits the characteristic shape that has been observed in similar devices [16,17,19], as predicted for charge-carrier recombination processes that follow the spin statistics of radical pairs [9,18,24] at low magnetic fields. Fig. 1(b) illustrates how the total static magnetic fields $B_{Net}$ experienced by each charge carrier in the pair is the vector sum of external field $B_0$ and $B_{hyp}$. Since the spatial localization of molecular orbitals of both charge carriers differs [25], the individual $B_{hyp}$ vectors are different in magnitude and direction. This difference leads to a distribution of the local magnetic fields and Zeeman energies, which inhibits degeneracy of the spin-pair states in the absence of an external field. As a field is applied, the spin-pair states shift in energy, causing the Zeeman splitting to either increase or decrease, depending on the mutual orientation of the external and the randomly oriented hyperfine field. Fig. 1(c-e) illustrates this process for decreasing Zeeman energies, which leads to a special case shown in Fig. 1(d), where the strength of $B_0$ is such that the levels with $S$ and $T_0$ character are aligned with $T_+$ and $T_-$, respectively. In this situation, spin-lattice relaxation processes ($T_1$), which usually involve energy exchange between the spin system and the lattice via phonons, become indistinguishable from mere decoherence processes ($T_2$), and the distinction between $T_1$ and $T_2$ is lost. As $T_1$ is typically larger than $T_2$, this indistinguishability implies that a strong dependence of the spin-relaxation time must exist within this magnetic-field regime.

For the pulsed EDMR experiment, well-established [26-30] pulse sequences for the determination of $T_2$ (i.e., electron spin-echo envelope modulation, ESEEM) and $T_1$ (i.e., inversion recovery) have been adapted from conventional inductively detected EPR spectroscopy by the addition of readout pulses [30], and the phase cycling sequence is modified appropriately [31]. While low-field continuous-wave EDMR experiments have been conducted previously on OLEDs [11, 32-36], SiC MOSFETS [37], and Si:P [38], experiments that involve pulsed EDMR at low excitation frequencies have only been demonstrated in the past in the context of spin-dependent processes in crystalline silicon systems [39,40]. Here, we report pEDMR on OLED recombination currents for frequencies between 40 MHz and 200 MHz, corresponding to resonance fields below 8 mT, where the effects of the random hyperfine fields on spin relaxation are expected to be most pronounced.

The OLED samples used in this study were fabricated on glass substrates (SPI supplies) with lithographically defined ITO contacts described earlier [16,27,41-43]. For the hole injection layer, $MoO_3$ (Sigma Aldrich) was spin-coated at 500 rpm for 3 seconds and 2000 rpm for 20 s, followed by thermal annealing at 110°C for 10 min., which forms a 20 nm thick layer [15]. SY-PPV [35,41,42] was dissolved in 1,2-dichlorobenzene at a concentration of 10 g/L, spin-coated inside a nitrogen-filled glovebox at

1000 rpm for 60 s, followed by thermal annealing at 110°C for 10 min. to form a 100 nm thick layer. Finally, 7 nm of calcium and 150 nm of aluminum were thermally evaporated at a pressure of $< 5 \times 10^{-6}$ mbar to form the electron injection layer and contact electrode, and the device was encapsulated using Araldite 2011 epoxy. The device structure is shown in Fig. S1(a), and a picture of the OLED with its 2 mm × 3 mm pixel under operating conditions is shown in Fig. S1(b) [45]. The I-V curve shown in Fig. S1(b), measured using a Keithley 2450 source meter confirms bipolar injection of electrons and holes. A detailed description of the pulsed EDMR experimental setup, device structure and the data acquisition can be found in the Supplemental Material [45].

Short rectangular RF pulses, as are required for coherent spin control [27-29,43,44,46], are challenging to produce. In conventional EPR spectrometers operating at microwave frequencies, this is achieved by pulse-forming units, which modulate a continuous-wave source with switches (e.g., PIN diodes) to form pulses of duration typically in the lower nanosecond range. At a much lower excitation frequency of ~100 MHz, the pulse length is on the order to the duration of a single oscillation period, and the finite response time of the RF switches could significantly distort the pulse shape. Longer, lower-power RF pulses are not desirable due to their reduced excitation bandwidth, which could potentially coherently manipulate only a fraction of the spectral lines. In addition, such soft pulses would require pulse sequences exceeding the charge-carrier spin coherence times. We therefore synthesized RF pulses directly using the AWG and confirm the excitation profile by calculating the Fourier transform of the pulse waveform. The excitation profile follows a *sinc* function (centered around the excitation frequency with a width that is inversely proportional to the pulse duration) that is characteristic of a rectangular pulse even for pulse lengths that are substantially lower than the oscillation period [45]. For multipulse sequences, the phases of the pulses are adjusted such that the oscillation is coherent throughout the entire sequence.

Figure 2(a) shows the EDMR signal following a 100 ns long 100 MHz RF pulse (RF power of 66 W) as a function of magnetic field as it is swept through resonance. The heatmap corresponds to the digitized transients of the OLED current change following the pulse, measured by sweeping the static magnetic field $B_0$ four times, whereas the solid white line represents the boxcar-integrated signal (the shaded region indicates the relative position of the integration interval) measured in a separate experiment (1 scan). In both cases, a sharp resonance is visible at $B_0 = 3.77$ mT, which corresponds to the g ≈ 2.00 resonance in SY-PPV [35,41,42]. The transient measurement has a single-scan signal-to-noise ratio of 10, whereas the single-scan signal-to-noise ratio of the integrated measurement is 22. In the following, we will only consider the boxcar-integrated measurements because of the increased sensitivity and the fact that the information contained in the temporal dynamics of the current response following the RF excitation is not particularly important for the present work.

In order to demonstrate coherent RF control, we perform transient nutation measurements, i.e., measurements of the integrated current following a 100 MHz RF pulse (66 W power) of varying duration [43,44,46] as a function of magnetic field swept across the resonance at $B_0 = 3.77$ mT. Results are shown in Fig. 2(b): the heatmap displays $\Delta Q$, the charge corresponding to the change in integrated current, as a function of pulse duration and magnetic field (bottom and left-hand axes), whereas the white line indicates $\Delta Q$ on resonance (right-hand axis). We observe several periods of oscillation in $\Delta Q$ on resonance, as marked by the dashed line. These oscillations quickly disappear when the field moves off-resonance. In order to test whether these oscillations are indeed Rabi oscillations [43], we perform this measurement at different RF powers between 20 W and 66 W as shown in Fig. 2(c). We indeed observe a scaling of the oscillation frequency $\Omega_R$ with the strength of $B_1$ as shown in the inset of Fig. 2(c). A spin beating of the signal due to the separation of the g-factors of the two spin species [43,44,46] is not resolved, however, presumably due to the comparatively quick decay of the oscillation envelope. Nevertheless, indications of an asymmetry of the transient integrated current profile with respect to the resonance field are apparent, as, e.g., at a pulse duration of approximately 100 ns in the map in panel (b).

In order to assess the spin-relaxation times $T_1$ and $T_2$ in the low-field regime, we performed electrically detected Hahn-echo [27-31] and inversion-recovery [26,29] measurements. The pulse sequences used in the experiments are shown in Fig. 3(a, d). These sequences correspond to the established methods for Hahn-echo [29,30,47] and inversion-recovery measurements [26,29] known from conventional EPR, with an additional π/2 readout pulse applied at the timing of the echo maximum. This readout pulse rotates the spins onto the axis of quantization to allow the charge signal $\Delta Q$ to be determined [48]. Non-resonant spin-independent signal contributions as well as additional echoes are removed by phase cycling (i.e., a 4-step phase cycling sequence for the Hahn echo and an 8-step sequence for inversion recovery) [27,31]. For each measurement, the pulse sequences depicted in Figs. 3(a, d) were repeated 2,000 times at a shot repetition rate of 500 Hz. In Fig. 3(b), the Hahn echo at $\nu = 100$ MHz, i.e., at resonance, is shown. For this measurement, the time interval $\tau$ [as defined in Fig. 3(a)] is varied, while $\tau'$ is kept fixed at $\tau' = 400$ ns. The duration of the π-pulse is kept constant at 62 ns for all measurements, and the RF power is determined by measuring the Rabi oscillations using Hahn-echo detection and adjusting the RF power to produce π-rotations. The delay between the readout-pulse and the start of the integration window is kept constant at 1500 ns, which yields the maximum single-shot signal-to-noise ratio of 2.8. The baseline before and after the leading and trailing edge of the echo scatters around zero, indicating complete cancellation of the non-resonant background signal by the correct choice of phase settings. Figure 3(c) shows the Hahn-echo amplitude as a function of $\tau$ from 88 ns to 1000 ns for $\tau = \tau'$, i.e., detected at the echo maximum, with all other parameters identical to Fig. 3(b), i.e., an electrically detected electron spin-echo envelope modulation (ESEEM) measurement. The electrically detected echo envelope follows a stretched exponential decay with a time constant of $T_2$. In Fig. 3(c) we obtain $T_2 = 469 \pm 7$ ns, which is comparable to the value previously obtained at 10 GHz [26].

In Figure 3(e-g), inversion-recovery echoes are shown as a function of $\tau$, with $\tau' = 400$ ns and $T = 200$ ns, 800 ns, and 50 µs [as defined in Fig. 3(d)], corresponding to the inverted echo, the echo that changes sign from negative to positive, and the fully recovered echo, respectively. This variation reflects the functional dependence of $\Delta Q$ on $T$, which represents an exponential decay from $-\Delta Q_0$ to $+\Delta Q_0$ with a time constant $T_1$. This dependence is also seen in Fig. 3(h), where $\Delta Q$ is shown as a function of $T$ with $\tau = \tau' = 400$ ns. Here, at 100 MHz, we obtain $T_1 = 1.8 \pm 0.1$ µs, which is considerably shorter than the value of $T_1$ previously obtained for SY-PPV at 10 GHz [26] and for similar conjugated polymer materials [28,49].

Finally, we explore the dependence of spin relaxation times $T_1$ and $T_2$ on excitation frequency below 200 MHz. Figure 4(a) shows the integrated current response following a single excitation pulse with a duration of 200 ns as a function of magnetic field and excitation frequency. Above 40 MHz, the dependence of the position of the respective resonance maxima is trivial and follows $g\mu_B B_0 = h\nu$, as indicated by the black dashed line. However, at excitation frequencies below 30 MHz, the resonance maxima start to deviate, as indicated by the solid white line, which is the result of a fit function that depends on the hyperfine field and net magnetic field experienced by the spins fitted to the experimentally determined resonance maxima. A detailed description of the fit function is given in Section G of the Supplemental Material [45]. This deviation from a linear function is caused by the competition of the externally applied magnetic field and the random hyperfine fields, which are comparable in magnitude, as well as the overlap with the excitation window of the EPR-inactive $B_1$ helicity, i.e., the Bloch-Siegert shift [33] which occurs when $B_1 \sim B_0$. A complete set of spectra is shown in Figure S7 of the Supplemental Material [45].

Figure 4(b) shows $T_1$ and $T_2$ values extracted from electrically detected inversion recovery and ESEEM measurements as discussed in Fig. 3, performed at different RF frequencies. The $T_1$ and $T_2$ values are plotted as a function of the resonance field, and the error bars indicate the uncertainty obtained from the numerical least-squares fit. The values for $T_2$ are quite constant and scatter around $T_2$=495±77 ns (i.e., within the error bars), which is consistent with $T_2$ values previously determined in SY-PPV at x-band frequencies [26]. In contrast, the $T_1$ values exhibit a distinct magnetic-field dependence: at frequencies above 60 MHz, $T_1$ is rather constant at 1.5±0.1 µs, whereas at lower frequencies, it decreases steeply to 0.9±0.2 µs at 44 MHz. Remarkably, for lower frequencies, $T_1$ again increases to 1.5±0.1 µs. This distinct behavior is consistent with the radical-pair model as illustrated in Fig. 1(c-e); in the intermediate field regime [cf. Fig. 1(d)], the $S$ and $T_0$ derived levels become energetically aligned with $T_+$ and $T_-$, respectively, making spin-lattice relaxation processes ($T_1$) indistinguishable from spin-spin relaxation processes ($T_2$). This indistinguishability is reflected by the results in Fig. 4(b), which indicate that $T_1 \approx T_2 = 0.9\pm0.2$ µs, whereas $T_1 > T_2$ for higher and lower magnetic fields, where this level alignment no longer occurs and $T_1$ processes require a phonon to exchange energy with the environment. This observation is qualitatively consistent with the ultrasmall magnetic field effect that is observed, e.g., in magnetoconductance [19].

We have demonstrated low-field room-temperature coherent spin control in OLEDs in a custom-made low-frequency EDMR spectrometer with direct RF pulse synthesis. The spin relaxation times $T_1$ and $T_2$ can be determined as a function of resonance frequency from 40-200 MHz through electrically detected ESEEM and inversion-recovery measurements. Our results indicate that $T_2$ remains relatively constant over the frequency range in question and is comparable to the $T_2$ value reported at higher excitation frequencies [26], whereas $T_1$, which is larger than $T_2$ for most frequencies, exhibits a distinct quenching around 44 MHz, below which it becomes comparable in magnitude to $T_2$. We attribute this dependency to a shift of the energy levels of the charge-carrier spin-pair states due to the random hyperfine fields experienced by each pair partner. These hyperfine fields become comparable to the external magnetic field at low frequencies and lead to an ultrasmall magnetic-field effect in observables depending on the radical-pair mechanism, including magnetoconductance in OLEDs [19].

**ACKNOWLEDGMENTS**

This work has been supported the U.S. Department of Energy, Office of Basic Energy Sciences, Division of Materials Sciences and Engineering under Award #DE-SC0000909. H. M. acknowledges support from the DFG (Germany) through the SFB 1277, project B03.

FIGURES

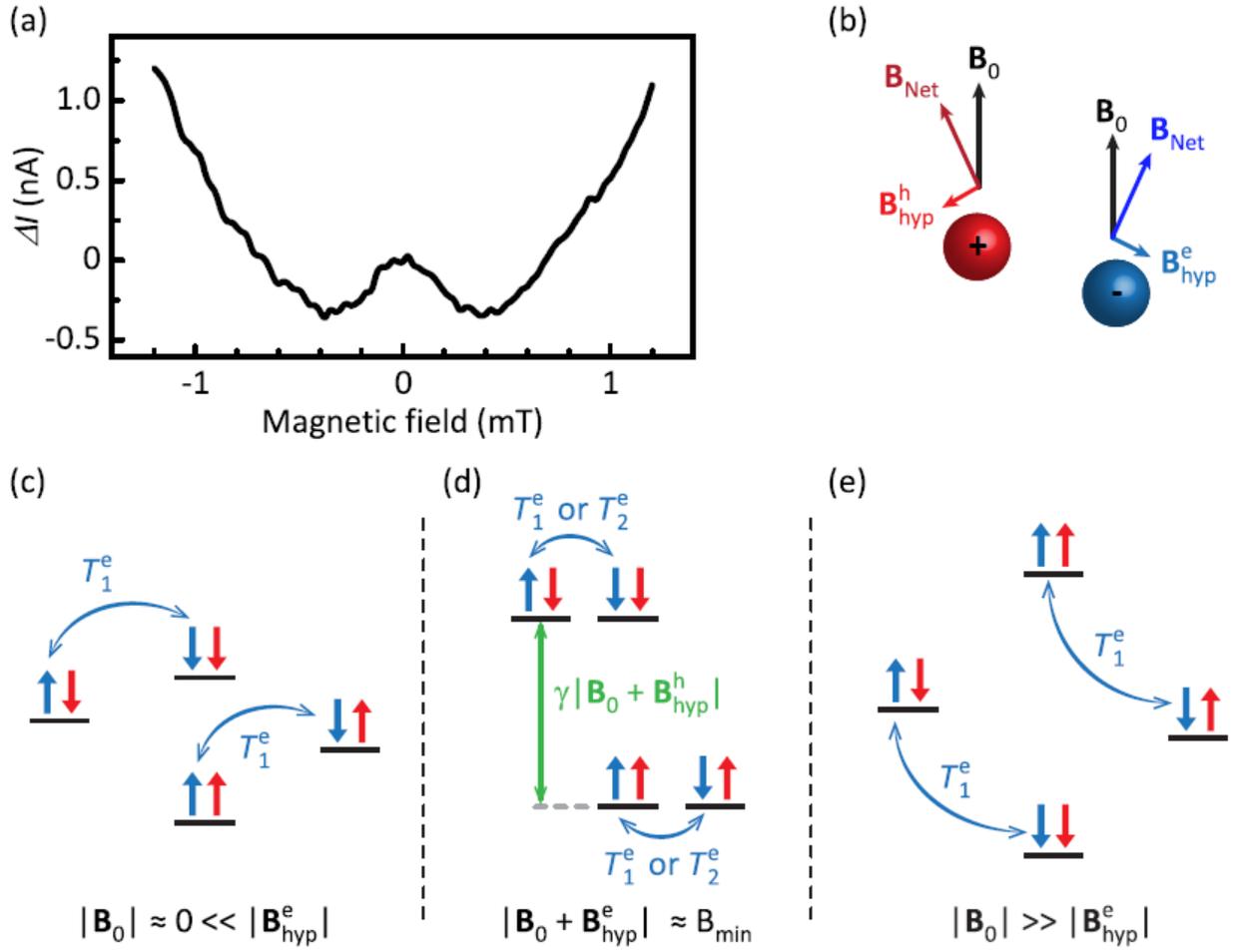

FIG. 1. (a) DC current change $\Delta I$ of a SY-PPV OLED as a function of an external magnetic field (magnetoconductance) below ~1.2 mT. The device has a 20 µA forward bias current. (b) Vector diagram of the total static magnetic fields of a charge-carrier pair. $B_0$ represents the external magnetic field, whereas $B_{hyp}^{e}$ and $B_{hyp}^{h}$ represent the randomly oriented hyperfine magnetic fields experienced by electron and hole, respectively. $B_{Net}$ is the vector sum of $B_0$ and the respective hyperfine fields. (c-e) Energy-level diagrams of two weakly spin-coupled charge-carrier pairs at (c) low magnetic field ($|B_0| \ll |B_{hyp}^{e}|$), (d) intermediate magnetic field ($|B_0 + B_{hyp}^{e}| \approx B_{min}$), and (e) high magnetic fields ($|B_0| \gg |B_{hyp}^{e}|$). Blue arrows indicate level transitions, i.e., spin-relaxation processes.

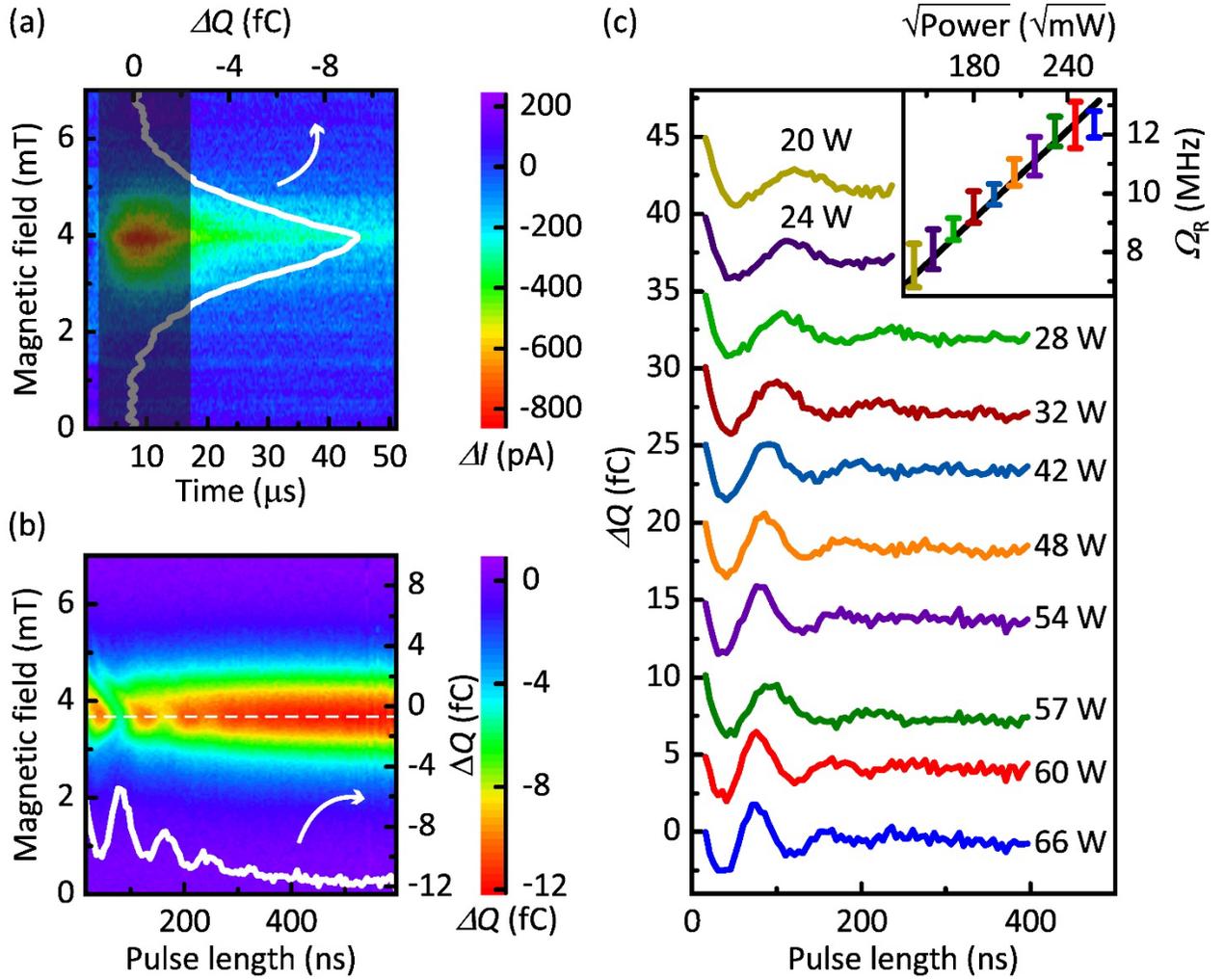

FIG. 2. Relative current changes ΔI following a short RF pulse of duration 100 ns and frequency of 100 MHz as a function of time after the pulse and magnetic field $B_0$. The white trace corresponds to ΔQ, the current change ΔI is integrated over a 15 μs interval (shaded region). For the integration a boxcar integrator is used. (b) ΔQ as a function of RF pulse length $τ$ and magnetic field. The white trace corresponds to ΔQ vs. $τ$ at the resonance maximum at $B_0$ =3.77 mT (dashed white line). (c) ΔQ as a function of $τ$ at the resonance maximum for various RF powers. The inset shows the oscillation frequency $Ω_R$, determined by Fourier transformation, as a function of RF power. The black line corresponds to a numerical least-squares fit of $Ω_R \propto \sqrt{\text{Power}}$.

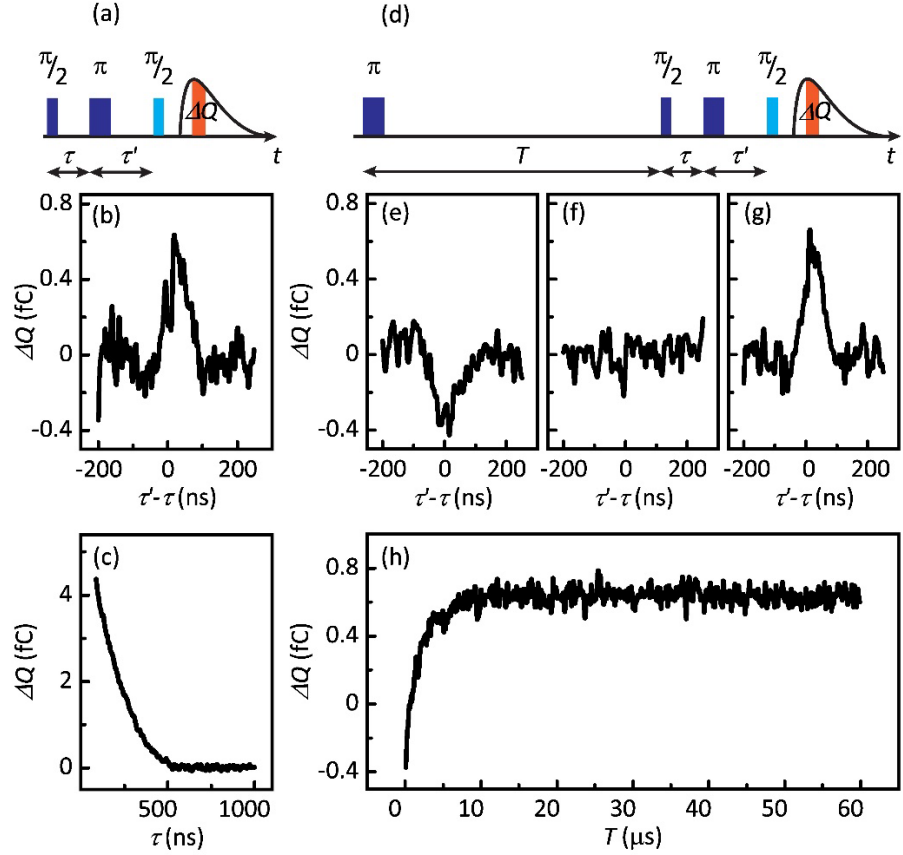

FIG. 3. (a) Pulse sequence for electrically detected Hahn echoes and ESEEM. (b) $\Delta Q$ as a function of $\tau$ with $\tau' = 400$ ns at room temperature and an excitation frequency of $\nu = 100$ MH, using a 4-step phase cycling sequence as described in Refs. 28 and 32 (i.e., an electrically detected Hahn echo). (c) $\Delta Q$ as a function of $\tau = \tau'$ (i.e., electrically detected ESEEM). (d) Pulse sequence for electrically detected inversion recovery. Here, $\tau = \tau'$ is kept constant and $T$ is varied. (e-g) $\Delta Q$ as a function of $\tau$ with $\tau' = 400$ ns and (e) $T=200$ ns, (f) $T = 800$ ns, and (g) $T=50$ μs using an 8-step phase cycling sequence (i.e., an electrically detected inversion echo). (h) $\Delta Q$ as a function of $T$ for $\tau = \tau'=400$ ns (i.e., electrically detected inversion recovery).

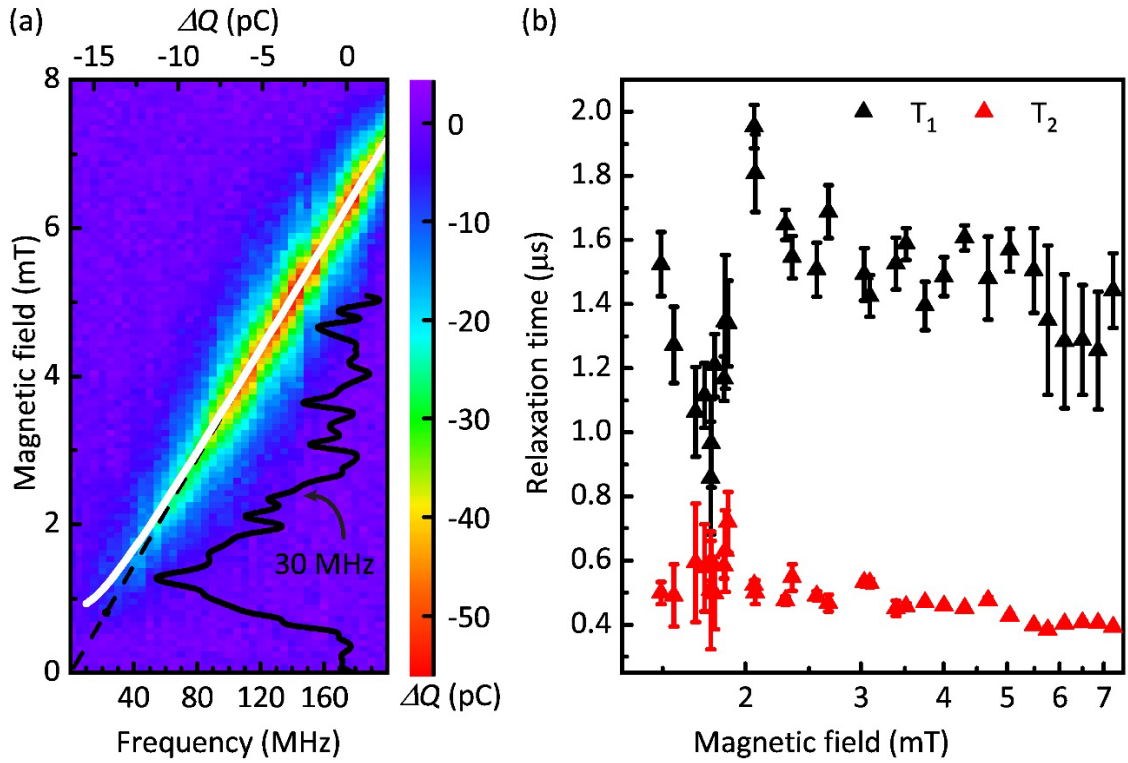

FIG. 4. (a) Integrated current change $\Delta Q$ following a 200 ns RF pulse as a function of excitation frequency and magnetic field. The solid white line corresponds to a fit function which depends on the net magnetic field experienced by spins to the resonance position of the measured spectra (see Section G in Ref. 45). The solid black line shows the spectrum measured at 30 MHz. The dashed black line indicates the resonance positions following the canonical relation $h\nu = g\mu_B B_0$. (b) Linear-log plot of spin relaxation times $T_1$ (black) and $T_2$ (red) as a function of magnetic field (i.e., the resonance center fields) measured at various excitation frequencies between 40 MHz and 200 MHz. Each data point represents the weighted average of several measurements under identical conditions.

**Spin relaxation dynamics of radical-pair processes at low magnetic fields**

# Supplemental Material

T. H. Tennahewa,[1] S. Hosseinzadeh,[1] S. I. Atwood,[1] H. Popli,[1] H. Malissa,[1,2] J. M. Lupton,[1,2] and C. Boehme[1]

[1]*Department of Physics and Astronomy, University of Utah, Salt Lake City, Utah 84112, USA*

[2]*Institut für Experimentelle und Angewandte Physik, Universität Regensburg, Universitätsstrasse 31, 93053 Regensburg, Germany*

# A) EXPERIMENTAL SETUP

The pulse EDMR setup is illustrated in Fig. S1(c) and consists of a homebuilt Helmholtz coil pair [35] driven by a computer-controlled current source (BK Precision 9202) that provides a static magnetic field $B_0$ of up to 15 mT, and a radio frequency (RF) coil (diameter 4 mm, length 4 mm, 4 number of turns, $L \approx 63.17$ nH) wrapped around the OLED in order to apply an oscillating field $B_1$ [1,2]. The RF pulses are directly synthesized by a computer-controlled arbitrary waveform generator (AWG, Wavepond Dax22000 with a sampling rate of 2.5 GHz) and amplified by two solid-state RF amplifiers (Mini-Circuits ZHL-100W-251-S+ with 46 dB gain in the frequency range 50 MHz to 250 MHz and ENI 5100L with 50 dB gain for frequencies <50 MHz). The pulses are applied to the RF coil and dissipated at a 50 Ω power resistor. In addition, the pulses are monitored through the voltage drop across the resistor using a potentiometer network [not shown in Fig. S1(c)]. The OLED is powered by a low-noise voltage source (SRS SIM928) that is adjusted to apply a DC device current of approximately 20 μA, and the device current change is measured using a transimpedance amplifier (SRS SR570 with a gain of 2 μA/V) with a 10 Hz to 100 kHz band-pass filter). The resulting voltage signal is again amplified (SRS SR560 with a gain of 50 and a 100 kHz low-pass filter) and is either directly recorded by a computer-controlled digitizer (AlazarTech ATS 9462) or, in the case of charge measurements, integrated over an interval of 15 μs using a boxcar integrator (SRS SR250 with a gain of 5 mV/V and a 10 Hz high-pass filter) [3], and the resulting charge signal is digitized by an analog-digital converter (ADC, National Instruments PCI 6251). The timing of the pulsed measurement is controlled through a pattern generator (Spincore Pulseblaster DDS-1-300) that triggers the AWG output, and, depending on the type of measurement, either the digitizer or the boxcar integrator and the ADC acquisition. Computer control of the field generator, the AWG, the pattern generator, and the digitizer/ADC is established through a self-written software using MATLAB.

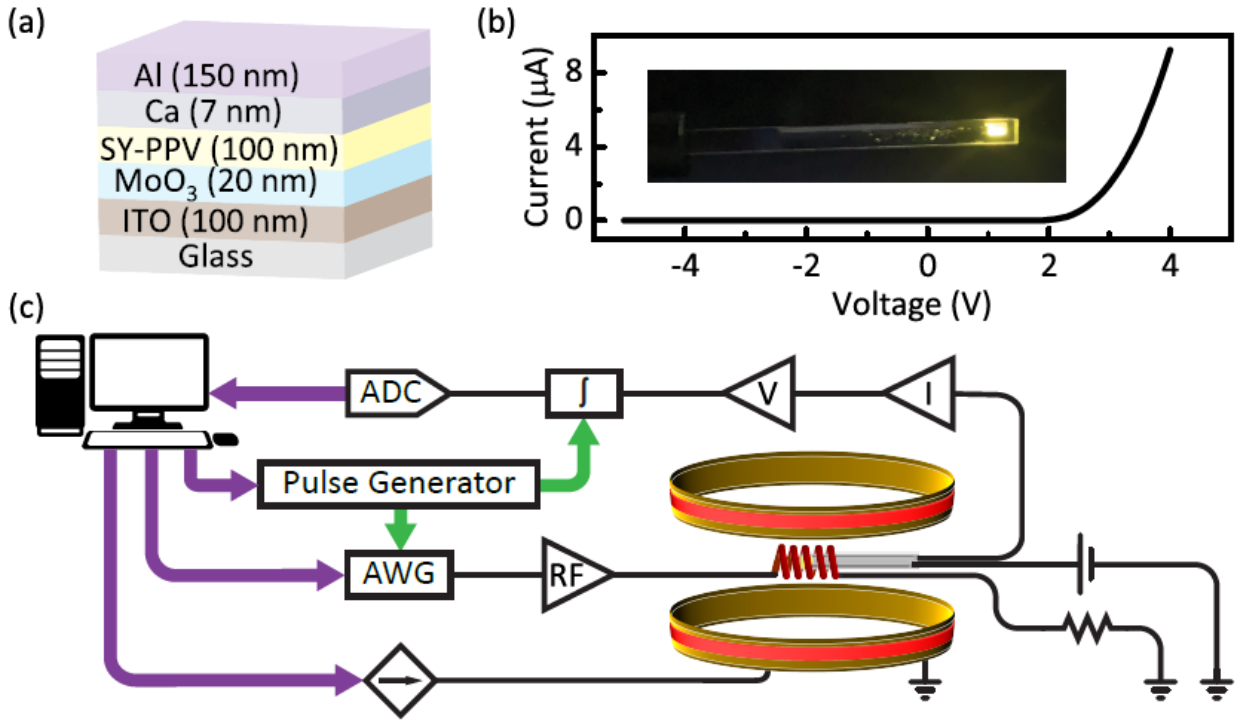

FIG. S1. (a) SY-PPV device structure used in this study. (b) I-V curve of a SY-PPV OLED at room temperature and a photograph of the OLED under bipolar charge-carrier injection. (c) Sketch of the experimental setup, consisting of (i) the circuit that applies a voltage bias to the OLED, and converts the current change $\Delta I$ to a voltage signal, (ii) the RF circuit that generates and amplifies the RF pulses at the coil, including a 50 $\Omega$ power resistor for dissipation, and (iii) the circuit that controls the magnetic field. Black lines correspond to analog signal lines, green arrows correspond to TTL trigger signals that control the timing of the experiment. Purple arrows correspond to digital control/acquisition lines.

## B) PULSE GENERATION

As described in the main text, we use Wavepond Dax22000 Arbitrary Waveform Generator (AWG) for pulse generation. It has a 2.5 GHz maximum sampling rate and analog outputs with 12-bit resolution. The AWG can be programmed to output any waveform. The AWG module's static random-access memory (SRAM) contains the user-defined waveform data—uploaded through a USB interface—and ultimately the AWG is put into the mode where the waveforms are synthesized when triggered [4].

The Fourier transform of a rectangular pulse of duration $\tau$ with constant amplitude ($A$) and the carrier frequency of the corresponding RF radiation ($f_0$) is a sinc function centered around the given $f_0$. In order to generate a pulse waveform with a duration that is shorter than the oscillation period, we calculate the real inverse Fourier transform as shown in Fig. S2.

$$f(t) = \frac{A\tau}{\pi} \int_{-\infty}^{\infty} \text{sinc}[\pi(f - f_0)\tau] \cos(2\pi f t) \, df$$

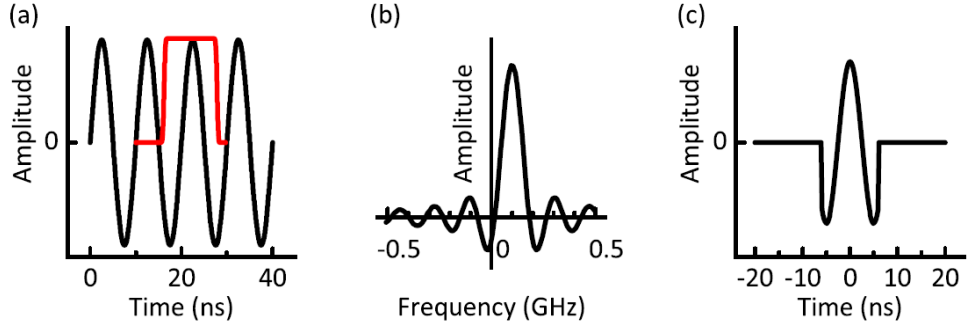

FIG. S2. Excitation pulse generation. (a) illustration of 12 ns pulse with 100 MHz RF frequency. (b) The excitation profile with the maximum peak is centered around 100 MHz (c) calculated pulse shape.

Evaluation of this expression gives a section of a rectangular pulse with a carrier frequency $f_0$ and duration $\tau$. In the following, we therefore generate our waveforms from sinusoidal functions with the appropriate frequency, phase, amplitude, and duration.

The function used to generate the pulse with this method for the magnetic field sweep is as follows. This function takes the amplitude `A`, RF frequency `RF`, and the number of points per excitation pulse `num_points` as inputs, and outputs the required waveform. The AWG requires the number of data points per programmed waveform to be modulo 16. This constraint is accounted for by the variable name `memory_depth`, which is adjusted depending on the pulse length required. The variable `awg_clk` signifies the sampling rate of the AWG and was fixed to be 2.5 GHz for all the experiments discussed here. Since the AWG has a 12-bit resolution (i.e., 0-4095 steps), the waveform is centered around step 2047.5.

```
step_size = 1/ awg_clk; % ns
num_points = pL/step_size;
t_array = 0:step_size:pL; % ns
y_array = zeros(size(t_array));

function Waveform = PulseGeneration(A,RF,num_points,t_array,y_array)
    for j = 1:length(t_array)
        y_array(j) = A* cos( 2*pi*RF*t_array(j));
```

```
        end
        if (mod(num_points,16) == 0)
            memory_depth = num_points;
        else
            memory_depth = num_points + (16 - mod(num_points,16));
        end
        Waveform = uint16(2047.5 + ones(1,memory_depth));
        Waveform(1:length(y_array))  =  uint16(ceil(2047.5  +  2047.5  *
        y_array));
end
```

The same approach was used in the Rabi nutation experiment. However, instead of having a single variable pulse length `pL`, an array of pulse lengths was fed to the above function. For the Hahn Echo experiment, the approach was slightly different. As mentioned in the main text, we used a two-pulse echo sequence [3] with fixed π/2 (`P0`) and π pulses (`P1`) and a subsequent π/2 projection pulse (`P9`) with 4-step phase cycling. The projection pulse is swept across the whole time range (`d0`). In the below code segment, `Pulse_phase[0]` and `Pulse_phase[pi]` represent the pulse in the +x/-x pulse channels used in phase cycling. `P_detec[+1]` and `P_detec[-1]` represent the corresponding summation or subtraction of the phase cycling step to the overall echo amplitude. The remaining variables and the MATLAB script are as follows.

```
P_detec = [ +1 -1 +1 -1];
Pulse_len = [P0 P1 P9] ;
Pulse_phase = [0 0 0 ; 0 0 pi ; pi 0 pi ; pi 0 0] ;
for x = 1: sy
    Echo_sum = 0 ;
    for phi = 1: size(Pulse_phase,1)
        pTrain = 2*d1 + d0(x) + P9;
        Pulse_pos = [0, d1, 2*d1+d0(x)] ;
        t_array = 0: step_size: pTrain ;
        y_array = zeros(size(t_array)) ;
        for j = 1: length(t_array)
            for p = 1: length(Pulse_pos)
                if (t_array(j) >= Pulse_pos(p) &
```

```
                            t_array(j) <= Pulse_pos(p) + Pulse_len(p))
                            y_array(j) = A * cos ( 2*pi* RF.*
                                t_array(j) + Pulse_phase(phi,p));
                    end
                end
            end
            if(mod( length(t_array),16) ==0)
                memory_depth = length(t_array);
            else
                memory_depth = length(t_array) +
                    (16 - mod( length(t_array),16));
            end
            Waveform = uint16(2047.5 + ones(1,memory_depth)) ;
            Waveform(1: length(y_array) )=uint16(
                ceil(2047.5 + 2047.5 * y_array));
```

At this point, the AWG is programmed to output the waveform, and once the trigger signal is set, data acquisition starts.

```
            Echo_sum = Echo_sum + P_detec(phi)*mean(AIvoltRead) ;
        end
        data_acc(1,x) = ((k-1) * data_acc(1,x) + mean(Echo_sum))/k ;
end
```

The MATLAB script for generating the inversion recovery sequence is as follows. In contrast to the Hahn-echo sequence, in the inversion recovery sequence, at the beginning of the sequence, there is a π (P2) pulse followed by the Hahn-echo sequence, and then at the end, a π/2 projection pulse (P9) with 8-step phase cycling.

```
Pulse_len = [P2 P0 P1 P9];
P_detec = [+1 -1 +1 -1 +1 -1 +1 -1];
Pulse_phase = [0 0 0 0; 0 0 0 pi; 0 pi 0 pi; 0 pi 0 0; pi 0 0 0;
    pi 0 0 pi; pi pi 0 pi ; pi pi 0 0] ;
for d = 1:sd
```

```
        Signal_sum = 0;
        for phi = 1: size(Pulse_phase,1)
            pTrain = d0 + 2*d1 + d2(d) + P9 ;
            Pulse_pos = [0 d2(d) d1+d2(d) d0+2*d1+d2(d)];
            t_array = 0: step_size: pTrain ;
            y_array = zeros(size(t_array)) ;
            for j = 1: length(t_array)
                for p = 1: length(Pulse_pos)
                    if (t_array(j) >= Pulse_pos(p) &
                            t_array(j) <= Pulse_pos(p) + Pulse_len(p))
                        y_array(j) = A * cos( 2*pi* RF.* t_array(j)
                                + Pulse_phase(phi,p)) ;
                    end
                end
            end
            if(mod( length(t_array),16) ==0)
                memory_depth = length(t_array);
            else
                memory_depth = length(t_array) +
                    (16 - mod( length(t_array),16));
            end
            Waveform = uint16(2047.5 + ones(1,memory_depth)) ;
            Waveform(1: length(y_array))=
                uint16(ceil(2047.5 + 2047.5 * y_array));
```

  At this point, the AWG is programmed to output the waveform, and once the trigger signal is set, data acquisition starts.

```
            Signal_sum = Signal_sum + P_detec(phi)*mean(AIvoltRead) ;
        end
        data_acc(1,d) = ((k-1) * data_acc(1,d) + mean(Signal_sum))/k ;
end
```

## C) DETECTION OF RABI OSCILLATIONS

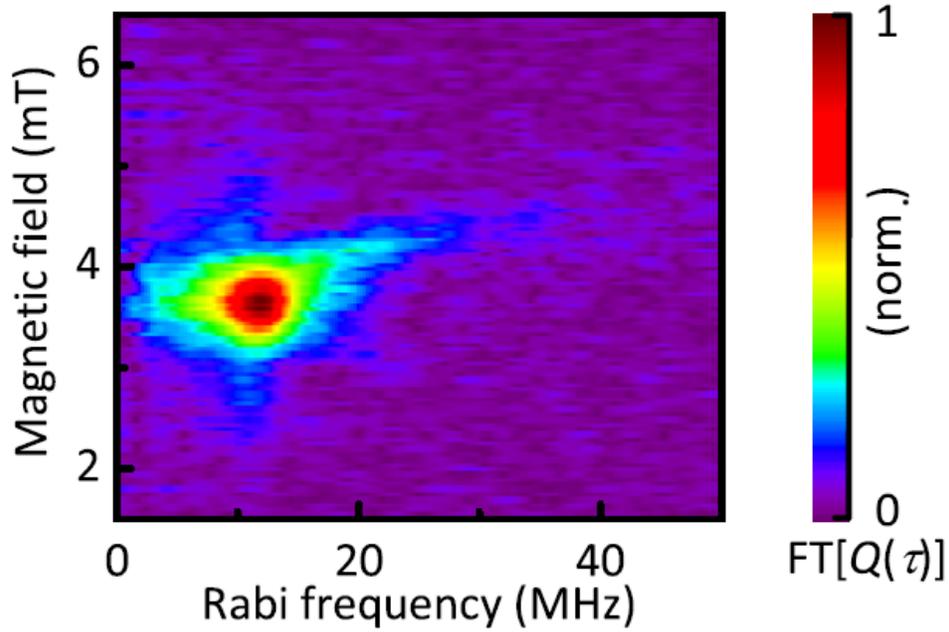

FIG. S3. Fourier transformation of the 2D spectrum of Rabi oscillations is shown in Fig. 2(b) of the main text. The FT Rabi oscillation spectrum does not appear to follow the fundamental Rabi nutation formula. In this low-frequency regime, the Rabi signal washes out due to the influence of hyperfine broadening. The data were recorded at room temperature on an OLED operated with a steady-state forward current of $I = 20$ μA, by measuring the integrated current transients after short-pulse RF excitation.

# D) DATA ACQUISITION

MATLAB is used to control the pulsed excitation and data acquisition processes. Here, a detailed explanation of the functioning of the experimental setup is given. The constant-bias OLED is placed at the center of the RF coil and the Helmholtz coils to provide an optimally homogeneous magnetic field. Using the MATLAB interface, experimental parameters such as the RF frequency and amplitude, pulse length, the number of shots, the shot repetition period, the number of points, and the range of the magnetic-field sweep are adjusted. The DLL libraries for the Dax22000 AWG, the BK precision 9202 current controllers, the DDS-1-300 pulse generator, and the NI PCI 6251 DAQ module are subsequently accessed for instrument control. For details on these libraries of the instruments, the reader is referred to the corresponding manuals [4,5]. We utilize two TTL trigger signals of the pulse generator. One is used to trigger the AWG output, and the other to trigger the signal integration on either the Boxcar integrator or acquisition on the digitizer.

After synthesizing the pulse in MATLAB, the waveform is uploaded to the AWG. Once the AWG is triggered, the pulse is applied to the coil. Shortly after the first trigger is applied, the second trigger activates the integrator data acquisition (or digitizer) and integrates the transient signal over the given time frame. For all our experiments, the integration window was set to 15 μs. Once the integration is complete, the boxcar integrator generates another TTL trigger signal. We use this signal to trigger the acquisition of the boxcar output with the ADC. One data point per *shot* (i.e., per actual pulse sequence) was collected for each magnetic field point, and the shot was repeated several times at the shot repetition rate. A moving-average approach was used to arrive at the final magnetic resonance spectrum. After identifying the magnetic resonance, the spectrum was used to find the resonance center and the initial position for the integration window for the remainder of the coherent-control experiments. The spectrum was fitted to a double Gaussian function, accounting for the distinct hyperfine-field distributions of electron and hole, to obtain the center of the resonance.

The position of the integration window was chosen so as to maximize the resulting EDMR signal; the underlying dynamics in the current transient are less relevant for our purpose and are lost in integration. Optimizing the integration window is crucial because it increases the overall signal amplitude. When selecting the integration window, ~85% of the signal should be inside it [see Fig. 2(a) in the main text]. Once the integration window and the resonance center are determined, the Rabi nutation experiment, the Hahn echo, and the inversion-recovery measurements can be performed.

Magnetoconductance measurements are carried out following Ref. [6].

# E) AMPLIFIER EFFECTS

We used two different power amplifiers for the experiments. To exclude any effects due to the different amplifiers, we measured the relaxation times for several frequencies using both amplifiers and compared the values of $T_1$ and $T_2$. Figure S4 shows two $T_1$ spectra measured at 40 MHz on two different OLEDs using both amplifiers in turn. Data were fitted with an exponential decay function resulting, for the ZHL-100W0251-S+ amplifier, in $T_1 = 1526.7 \pm 108.9$ ns, and for the ENI 5100L amplifier in $T_1 = 1513.1 \pm 265.1$ ns. The data demonstrate that the results shown in the text are independent of the amplifier and the OLED devices.

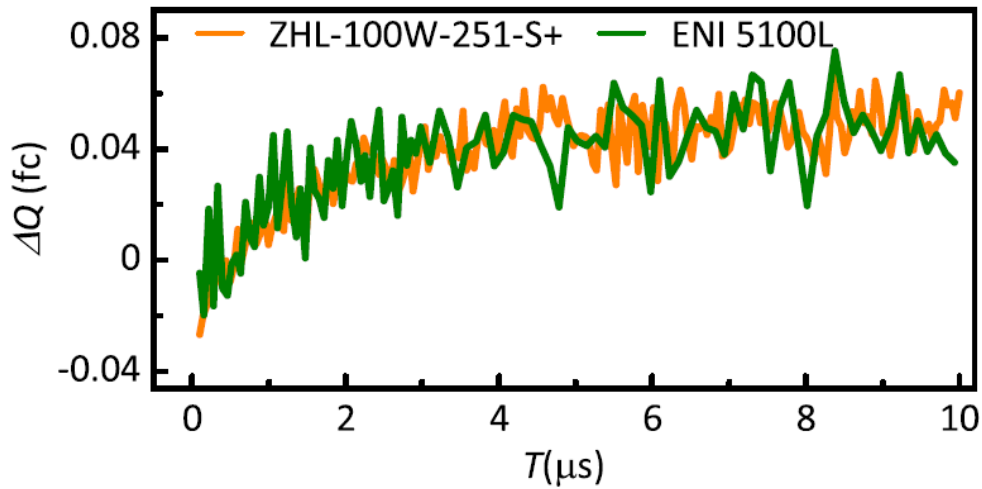

FIG. S4. Inversion-recovery spectra measured at 40 MHz on two OLEDs using two different power amplifiers.

## F) SPIN-RELAXATION TIME MEASUREMENTS

Figure 4(b) in the main text represents the Linear-log plot of spin relaxation times as a function of the resonance center fields measured at various excitation frequencies between 40 MHz and 200 MHz. For frequencies below 60 MHz, the experimental resonance centers have deviated from the magnetic field centers given by $h\upsilon/g\mu_B$. This deviation is caused by the competition of the externally applied magnetic field and the randomly oriented hyperfine fields in the environment. Figure S5 shows weighted average $T_1$ and $T_2$ spin relaxation times as a function of the theoretical magnetic field centers and Fig. S6 represents the summary of all the measurements.

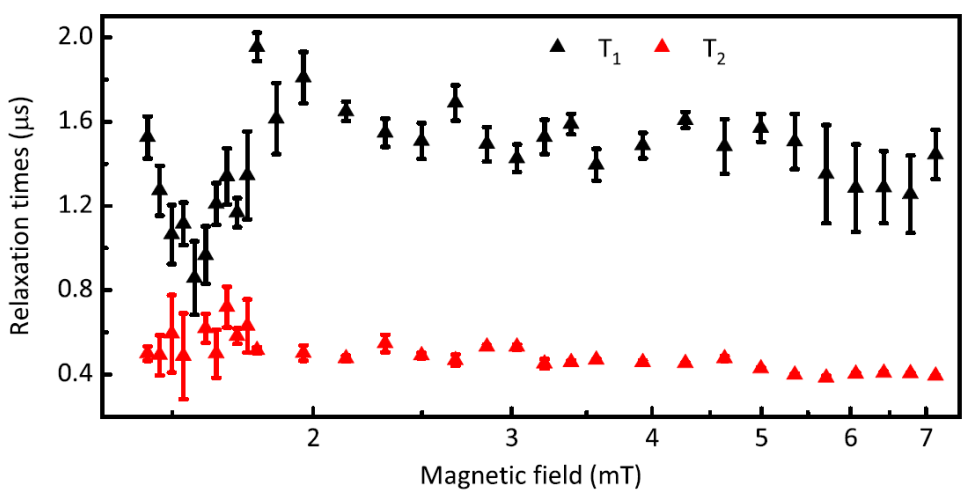

FIG. S5. Weighted average of $T_1$ and $T_2$ spin-relaxation times measured from 40 MHz (1.4 mT) to 200 MHz (7.1 mT) as a function of magnetic field centers given by $h\upsilon/g\mu_B$. The spectra between 50 MHz and 200 MHz were measured using the ZHL-100W0251-S+ power amplifier, and the spectra below 50 MHz were measured using the ENI 5100L power amplifier.

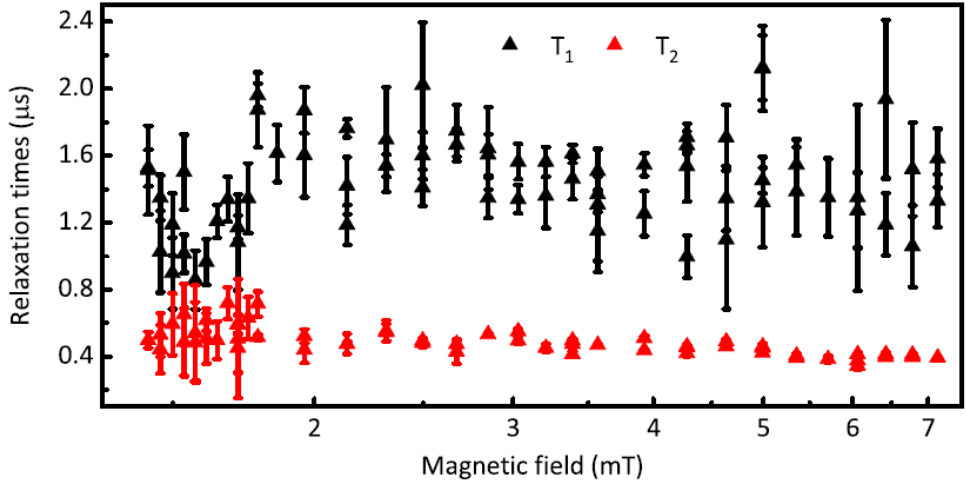

FIG. S6. Summary of all the $T_1$ and $T_2$ spin relaxation times measured from 40 MHz (1.4 mT) to 200 MHz (7.1 mT) as a function of magnetic field centers given by $h\nu/g\mu_B$. The spectra between 50 MHz and 200 MHz were measured using the ZHL-100W0251-S+ power amplifier, and the spectra below 50 MHz were measured using the ENI 5100L power amplifier.

## G) INFLUENCE OF THE HYPERFINE INTERACTION FIELDS

The net magnetic field experienced by the spins in the ensemble depends on the applied static magnetic field and the random hyperfine fields in the ensemble. As shown in the main text, due to the comparable magnitudes of random hyperfine fields ($B_{\text{hyp}}$) and $B_0$ fields, the resonance centers deviated from the relation $g\mu_B B_0 = h\nu$. In order to find this deviation, the data were fitted to the relation

$$B_{\text{Net}} = \sqrt{B_{\text{hyp}}^2 + \left(\frac{h\nu}{g\mu_B}\right)^2}.$$

To find the resonance centers, spectra were fitted with a combination of a double Gaussian function and a first order polynomial which takes the magnetoresistance effects into consideration. Then the minima of the spectra were determined, and these were fitted with the above equation. The resulting value of $B_{\text{hyp}} = 0.9$ mT is similar to previously obtained results [7].

For low frequencies, some of the individual magnetic resonance spectra plotted in Fig. 4(a) are displayed in Fig. S7. There is a clear indication that the resonance centers saturate below 40 MHz.

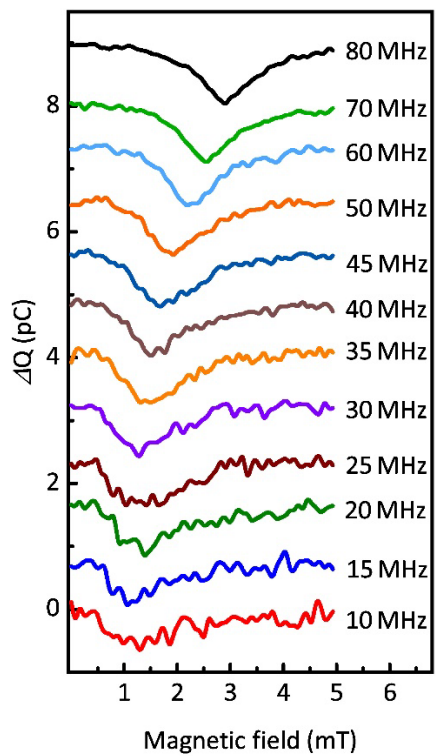

FIG. S7. Integrated current change $\Delta Q$ following a 200 ns RF pulse as a function of magnetic field measured at different excitation frequencies. The resonance centers begin to deviate from the linear behavior below 40 MHz and the center field values saturate.